\documentclass[twocolumn]{aastex62}
\shorttitle{A New Detection of Extragalactic AME in NGC\,4725}
\shortauthors{MURPHY ET AL.}
\accepted{to ApJ; May 15, 2018}

\begin{document}

\title{A New Detection of Extragalactic Anomalous Microwave Emission in a Compact, Optically-Faint Region of NGC\,4725}

\author{E.J.\,Murphy}
\email{emurphy@nrao.edu}
\affiliation{National Radio Astronomy Observatory, 520 Edgemont Road, Charlottesville, VA 22903, USA}

\author{S.T.\,Linden}
\affiliation{Department of Astronomy, University of Virginia, 530 McCormick Road,Charlottesville, VA 22904, USA}

\author{D.\,Dong}
\affiliation{California Institute of Technology, MC 100-22, Pasadena, CA 91125, USA}

\author{B.S.\,Hensley}
\affiliation{Jet Propulsion Laboratory, California Institute of Technology, 4800 Oak Grove Drive, Pasadena, CA 91109, USA}

\author{E.\,Momjian}
\affiliation{National Radio Astronomy Observatory, P.O. Box O, 1003 Lopezville Road, Socorro, NM 87801, USA}

\author{G.\,Helou}
\affiliation{Infrared Processing and Analysis Center, California Institute of Technology, MC 220-6, Pasadena CA, 91125, USA}

\author{A.S.\,Evans}
\affiliation{Department of Astronomy, University of Virginia, 3530 McCormick Road,Charlottesville, VA 22904, USA}
\affiliation{National Radio Astronomy Observatory, 520 Edgemont Road, Charlottesville, VA 22903, USA}

\begin{abstract}
We discuss the nature of a discrete, compact radio source (NGC\,4725\,B) located $\approx$1.9\,kpc from the nucleus in the nearby star-forming galaxy NGC\,4725, which we believe to be a new detection of extragalactic Anomalous Microwave Emission (AME).  
Based on detections at 3, 15, 22, 33, and 44\,GHz, NGC\,4725\,B is a $\mu$Jy radio source peaking at $\approx$33\,GHz. While the source is not identified in $BVRI$ photometry, we detect counterparts in the mid-infrared {\it Spitzer}/IRAC bands (3.6, 4.5, 5.8, 8.0\,$\mu$m) that appear to be associated with dust emission in the central region of NGC\,4725. Consequently, we conclude that NGC\,4725\,B is a new detection of AME, and very similar to a recent detection of AME in an outer-disk star-forming region in NGC\,6946. We find that models of electric dipole emission from rapidly rotating ultra-small grains are able to reproduce the radio spectrum for reasonable interstellar medium conditions. Given the lack of an optical counterpart and the shape of the radio spectrum, NGC\,4725\,B appears consistent with a nascent star-forming region in which young ($\lesssim 3$\,Myr) massive stars are still highly enshrouded by their natal cocoons of gas and dust with insufficient supernovae occurring to produce a measurable amount of synchrotron emission.
\end{abstract}
\keywords{dust, extinction -- galaxies: individual (NGC\,4725) -- H{\sc ii} regions -- radio continuum: general -- stars: formation}

\section{Introduction}
Radio continuum (i.e., $1 \lesssim \nu \lesssim 100$\,GHz) emission from galaxies is thought to be primarily composed of two components associated with massive (i.e., $\gtrsim 8M_{\sun}$) star formation: 
free-free emission, which is directly related to the production rate of ionizing (Lyman continuum) photons in H{\sc ii} regions, and non-thermal emission which arises from cosmic-ray electrons/positrons propagating through the magnetized interstellar medium (ISM) after having been accelerated by supernova (SN) remnants \citep[e.g.,][]{kk95}.   
The non-thermal emission component typically has a steep spectrum ($S_{\nu} \propto \nu^{\alpha}$, where $\alpha \sim -0.8$), while the free-free component is relatively flat \citep[$\alpha\sim-0.1$; e.g., ][]{jc92}.  

An often overlooked additional component, that can emit as strongly as these other two, is anomalous microwave emission \citep[AME; for a recent review, see][]{AME18}.  
First discovered through accurate separation of Galactic foreground components in cosmic microwave background experiments, AME was found to contribute much of the radio continuum from $\sim10-90$\,GHz while having strong correlation with 100\,$\micron$ thermal emission from interstellar dust \citep[e.g.][]{ak96b,eml97}.  

The leading explanation for AME is rotational emission from ultra-small ($a \la 1$\,nm) dust grains (i.e., ``spinning dust"), which was first postulated by \citet{wce57}.  
In this model, rapidly rotating very small grains with non-zero electric dipole moments produce the observed microwave emission \citep{dl98b,plsd11,hd17}.  
Magnetic dipole radiation from thermal fluctuations in the magnetization of interstellar dust grains \citep{dl99,dh13}, may also be contributing to observed AME, particularly at higher frequencies ($\gtrsim$50\,GHz).  

Since its initial discovery, AME has been observed by a number of experiments and in a variety of environments.    
The majority of AME detections have been reported for regions in the Galaxy, with surprisingly few extragalactic detections in the literature \citep[e.g.,][]{ejm10,as10,fpi10,cbot10,bh15,Planck2015-XXV}.  
The first detection of AME outside of the Galaxy was for a single star-forming region (Enuc.\,4) in the disk of the nearby galaxy NGC\,6946 \citep{ejm10}, where a spinning dust model for the spectrum was preferred over a free-free-only model.  
The detection was confirmed by follow-up observations with the Arcminute Microkelvin Imager telescope at 15 -- 18\,GHz \citep{as10} and the Combined Array for Research in Millimeter-wave Astronomy at 30 -- 100\,GHz \citep{bh15}.

In this Paper we present a new likely detection of AME for a discrete region located within the disk of the nearby galaxy NGC\,4725.  
This source (NGC\,4725\,B) was first discovered by 33\,GHz Karl G. Jansky Very Large Array (VLA) observations carried out as part of the Star Formation in Radio Survey \citep[SFRS;][]{ejm18a}.  
This Paper is organized as follows.  
The data and analysis procedures are presented in \S2.  
The results are then presented in \S3 and discussed in \S4.  
A brief summary of our findings are given in \S5.

\section{Data and Analysis}
\begin{figure}[t!]
\epsscale{1.2}
\plotone{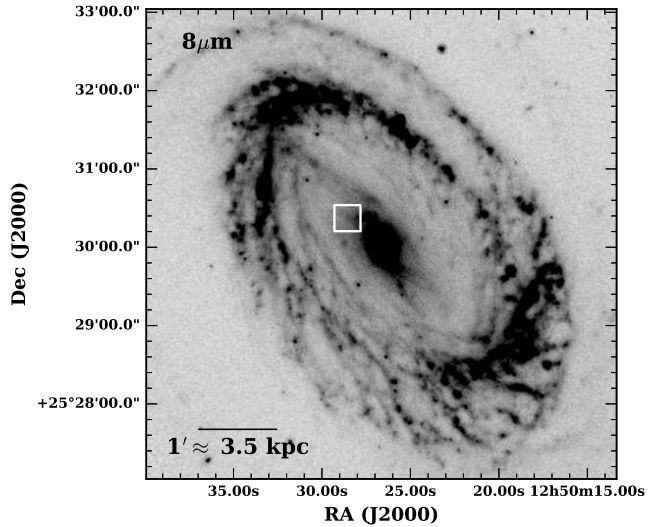}
\caption{
A {\it Spitzer} 8\,$\mu$m grayscale image of NGC\,4725 observed as part of SINGS \citep{rck03}.  
The $15\arcsec \times 15\arcsec$ box near the nucleus highlights the location of NGC\,4725\,B and indicates the size of the panels shown in Figures \ref{fig:radpan} and \ref{fig:ancpan}.  
A scale bar is shown in the bottom left corner of the image.  
}
\label{fig:gal}
\end{figure}

NGC\,4725 is a nearby \citep[$d_{L} = 11.9$\,Mpc][]{hkp01} barred ringed spiral galaxy that hosts an AGN (Sy2) nucleus \citep{jm10}.  
Located at J2000 $\alpha =12^\mathrm{h}50^\mathrm{m}28\fs48, \delta =+25\degr30\arcmin22\farcs5$, NGC\,4725\,B was detected at 33\,GHz as part of the SFRS \cite[][]{ejm18a}, comprising nuclear and extranuclear star-forming regions in 56 nearby galaxies ($d_{L} < 30$\,Mpc) observed as part of the {\it Spitzer} Infrared Nearby Galaxies Survey \citep[SINGS;][]{rck03} and Key Insights on Nearby Galaxies: a Far-Infrared Survey with {\it Herschel} \citep[KINGFISH;][]{kf11} legacy programs.
In Figure \ref{fig:gal}, a {\it Spitzer} 8\,$\mu$m image of NGC\,4725 is shown, highlighting the location of NGC\,4725\,B at a galactocentric radius of $\approx$1.9\,kpc by a $15\arcsec \times15\arcsec$ ($\approx 900\,{\rm pc} \times 900$\,pc) box.   

\subsection{Radio Data}
Radio data were obtained as part of multiple campaigns using the VLA, which are summarized in Table \ref{tbl-1}.  
Initial observations in the Ka-band ($26.5-40$\,GHz) were taken in the D-configuration 2013 March as part of VLA/13A-129.  
For these observations the 3-bit samplers were used, yielding 8\,GHz of instantaneous bandwidth in  2\,GHz wide basebands centered at 30, 32, 34, and 36\,GHz. 
Details on the observing strategy along with the reduction and imaging procedure can be found in \citet{ejm18a}.  

Additional data in the S- ($2-4$\,GHz) and Ku- ($12-18$\,GHz) bands were observed as part of VLA/13B-215. We refer to these observations and their images as 3 and 15\,GHz, respectively.  
Observations at 3\,GHz were taken 2013 November in the B-configuration using the 8-bit samplers while 15\,GHz observations were taken 2014 November in the C-configuration using the 3-bit samplers. The choice of configurations for the 3 and 15\,GHz observations were to most closely match the angular resolution (i.e., FWHM of the synthesized beam $\theta_{1/2} \approx 2\arcsec$) of the 33\,GHz data.  

Upon discovering the peculiar nature of NGC\,4725\,B \citep{ejm18a}, we successfully applied for Director’s Discretionary Time (DDT) to obtain new observations in the K- ($18-26.5$) and Q- ($40-50$\,GHz) bands,
along with an additional observation at 33\,GHz to look for possible variability. 
The K- and Q-band observations also utilized the 3-bit samplers, providing 8\,GHz of frequency span in each receiver band centered at 22 and 44\,GHz, respectively. 
We refer to these results as the 22 and 44\,GHz images.
These data were taken 2017 July in the C-configuration, and as such, the angular resolution of the DDT data spans a large range (see Table \ref{tbl-1}).  

Detailed information on the observing strategy and calibration procedure for the 3 and 15\,GHz data can be found in E. J. Murphy et al. (in preparation, 2018), but closely follow those for the 33\,GHz data as described in \citep{ejm18a}.  
To summarize, we followed standard calibration procedures, using the VLA calibration pipeline built on the Common Astronomy Software Applications \citep[CASA;][]{casa} versions 4.6.0 and 4.7.0. After each initial pipeline run, we manually inspected the calibration tables and visibilities for signs of instrumental problems (e.g., bad deformatters) and RFI, flagging correspondingly. After flagging, we re-ran the pipeline, and repeated this process until we could not detect any further signs of bad data.  

\begin{figure*}[t!]
\epsscale{1.1}
\plotone{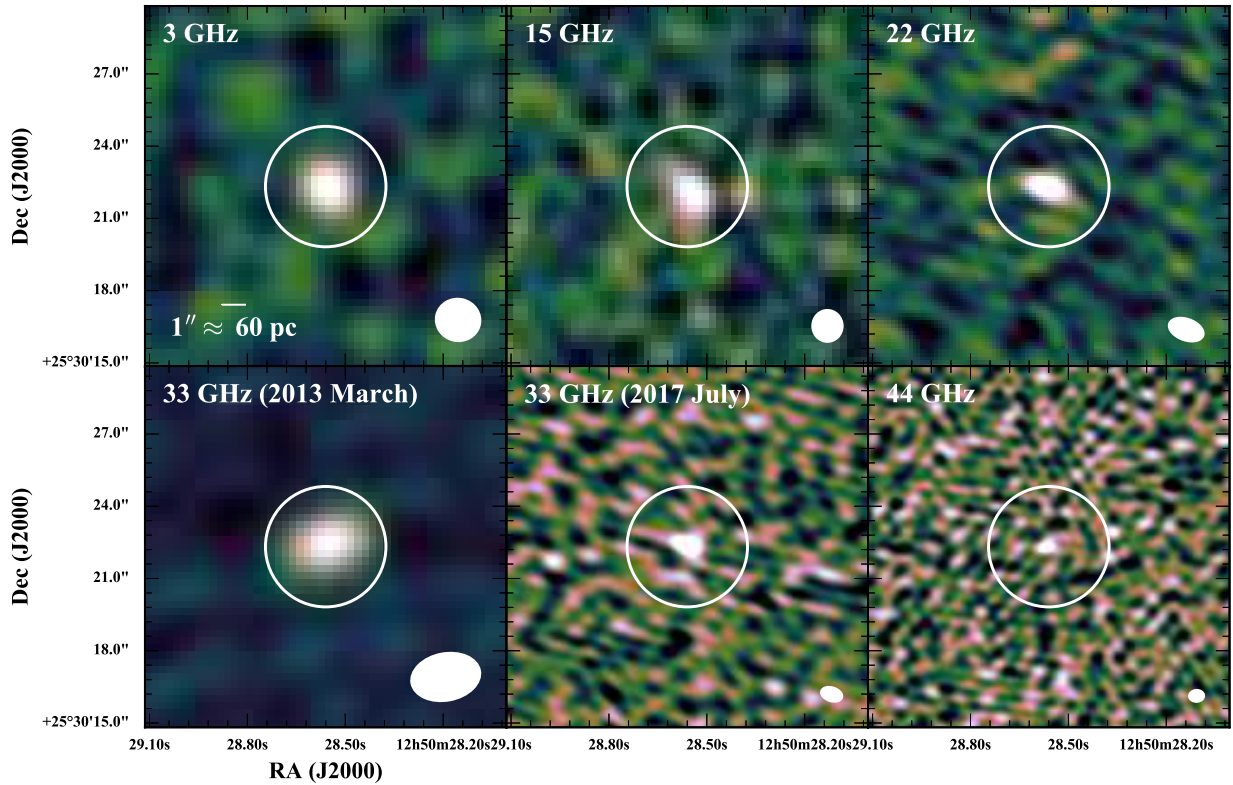}
\caption{
Colorscale images displayed with a linear scaling at all observed radio frequencies (i.e., 3, 15, 22, 33, and 44\,GHz), showing that NGC\,4725\,B is clearly detected in all bands.  
Images for the original and re-observed 33\,GHz data, taken in different configurations, are shown.  
Each panel is $15\arcsec \times 15\arcsec$, which is illustrated in Figure \ref{fig:gal}.  
The circle is centered on the source in each panel, and has a diameter of 5\arcsec ($\approx 300$\,pc); this is also the case in Figure \ref{fig:ancpan} for comparison purposes.  
The size and orientation of the synthesized beams are illustrated in the bottom right corner of each panel.  A scale bar is shown in the bottom left corner of the first panel.  
}
\label{fig:radpan}
\end{figure*}

As with the data calibration, a detailed description of the imaging procedure used here can be found in \citet{ejm18a}.  
To summarize, calibrated VLA measurement sets for each source were imaged using the task {\sc tclean} in CASA version 4.7.0. The mode of {\sc tclean} was set to multi-frequency synthesis \citep[{\sc mfs};][]{mfs1,mfs2}. We chose to use {\it Briggs} weighting with {\sc robust=0.5}, and set the variable {\sc nterms=2}, which allows the cleaning procedure to also model the spectral index variations on the sky. We additionally made use of the multiscale clean option \citep{msclean,msmfs} to help deconvolve extended low-intensity emission, searching for structures with scales $\approx$1 and 3 times the FWHM of the synthesized beam. The only exception to this basic strategy is at Q-band data, for which we set the variable {\sc nterms=1} and did not make use of the multiscale clean option as this did not seem warranted due to the source being low signal-to-noise at this frequency. 
The synthesized beam, point source rms, and surface brightness rms for each final image is given in Table \ref{tbl-1}. 
The largest angular scale (LAS) that our imaging is sensitive to across all frequency bands and antenna configurations is $\theta_{\rm LAS} \approx 32\arcsec - 97\arcsec$.  
In Figure \ref{fig:radpan}, $15\arcsec \times 15\arcsec$ ($\approx 900\,{\rm pc} \times 900$\,pc) cutouts centered on the location of NGC\,4725\,B are shown for all radio data.



\begin{deluxetable}{c|ccccc}
\tablecaption{Observational Data \label{tbl-1}}
\tabletypesize{\scriptsize}
\tablewidth{0pt}
\tablehead{
\colhead{Freq.}  & \colhead{Obs. Date} & \colhead{Synthesized}& \colhead{$\sigma$}& \colhead{$\sigma_{T_{b}}$}& \colhead{$S_{*}^{\dagger}$}\\
\colhead{(GHz)}  & \colhead{} & \colhead{Beam}& \colhead{($\mu$Jy\,bm$^{-1}$)}& \colhead{(mK)}& \colhead{($\mu$Jy)}
}
\startdata
  3  & 2013-11-04  &  $1\farcs85 \times 1\farcs76$  & 13.7  &  569.  & $  152.\pm   11.$  \\
 15  & 2014-11-11  &  $1\farcs33 \times 1\farcs26$  & 9.03  &  29.2  & $  217.\pm   28.$  \\
 22  & 2017-07-03  &  $1\farcs50 \times 0\farcs89$  & 20.5  &  38.3  & $  312.\pm   26.$  \\
 33  & 2013-03-07  &  $2\farcs89 \times 1\farcs97$  & 10.7  &  2.09  & $  396.\pm   11.$  \\
 33  & 2017-07-03  &  $0\farcs91 \times 0\farcs58$  & 21.4  &  45.3  & $  401.\pm   24.$  \\
 44  & 2017-07-03  &  $0\farcs63 \times 0\farcs50$  & 32.9  &  66.7  & $  279.\pm   56.$    
\enddata
\tablenotetext{\dagger}{Photometry measured after convolving all data to match the beam of the 2013 March 33\,GHz data.}
\end{deluxetable}

\subsection{Ancillary Data}
\begin{figure*}[t!]
\epsscale{1.1}
\plotone{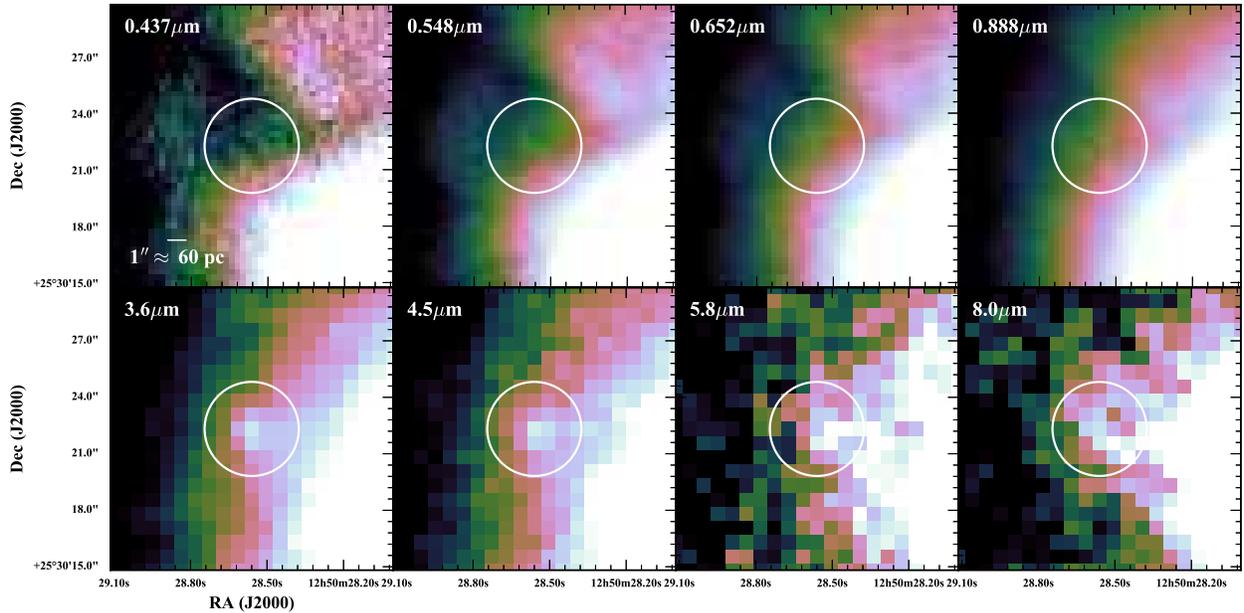}
\caption{
Colorscale optical ($BVRI$) and {\it Spitzer}/IRAC (3.6, 4.5, 5.8, and 8\,$\mu$m) images observed as part of SINGS \citep{rck03}, each displayed with a logarithmic scaling.  
As is the case for Figure \ref{fig:radpan}, each panel is $15\arcsec \times 15\arcsec$, which is illustrated in Figure \ref{fig:gal}.  
The circle is centered on the source in each panel, and has a diameter of 5\arcsec ($\approx 300$\,pc); this is also the case in Figure \ref{fig:radpan} for comparison purposes.  
A scale bar is shown in the bottom left corner of the first panel.  
While there is no clear detection of the radio source (NGC\,4725\,B) in the optical bands, there is an apparent coincident feature in all IRAC bands.
}
\label{fig:ancpan}
\end{figure*}

We make use of archival optical $BVRI$ and {\it Spitzer}/IRAC 3.6, 4.5, 5.8, and 8.0\,$\mu$m data obtained by the SINGS legacy science program \citep{rck03} to search for potential counterparts to the radio emission of NGC\,4725\,B. 
The seeing-limited optical imaging was observed using the Kitt Peak National Observatory 2.1\,m telescope and achieves angular resolutions of $\sim 1-2\arcsec$.  
The {\it Spitzer}/IRAC imaging sensitivities are $0.02-0.12$\,MJy\,sr$^{-1}$ with angular resolutions comparable to the $BVRI$ imaging. 
Consequently, these data are sufficient to map the stellar emission and mid-infrared emitting dust over the entire optical extent of NGC 4725.  
Similar to Figure \ref{fig:radpan}, $15\arcsec \times 15\arcsec$ ($\approx 900\,{\rm pc} \times 900$\,pc) cutouts centered on the location of NGC\,4725\,B are shown in Figure \ref{fig:ancpan} at each of these wavelengths.

\subsection{Radio Photometry \label{sec:radphot}}
Photometry was carried out at all frequencies before applying a primary beam correction. 
To measure the flux densities of the source at each frequency, we first beam-match the data to the lowest resolution beam among all images (i.e., the 2013 March 33\,GHz data) using the CASA task {\sc imsmooth}. 
Next, the CASA task {\sc imfit} was used to fit an elliptical Gaussian to the emission within a circular aperture having a radius equal to the FWHM of the synthesized beam major axis.  
A primary beam correction was then applied to the reported peak brightnesses and integrated flux densities (along with their errors) using the appropriate frequency-dependent primary beam correction given in EVLA Memo\#\,195 (Perley 2016)\footnote{\url{http://library.nrao.edu/public/memos/evla/EVLAM_195.pdf}}.    
Primary beam corrections were $<10\%$ for all cases except the original (2013 March) 33\,GHz observations, for which the phase center was located at the galaxy nucleus, leading to a primary beam correction factor of 1.52.  

The source is unresolved in the native resolution images using the criterion that the fitted major axis $\phi_{\rm M}$ be at least $2\sigma_{\phi_{\rm M}}$ larger than the FWHM of the synthesized beam major axis.  
Similarly, the source appears unresolved in the beam-matched images at all frequencies, except at 15\,GHz ($2.3\sigma$ significance).  
Consequently, we take the total flux density ($S_{*}$) for the unresolved cases to be the geometric mean of the peak brightness and integrated flux density reported by {\sc imfit}.  
We choose this value, as it provides the most accurate measurement for the flux density of a weak source, given that the uncertainties on the peak brightness and integrated flux density from the elliptical Gaussian fitting are anti-correlated \citep{jc97}.  
At 15\,GHz, the integrated flux density from the fit is taken as $S_{*}$.  
The total flux densities, along with their uncertainties, are given in Table \ref{tbl-1} at each frequency.    
We note that the 33\,GHz flux densities measured at the two separate epochs are entirely consistent within their uncertainties. 
The radio spectrum of NGC\,4725\,B is shown in Figure \ref{fig:spec}.  

While {\sc imfit} suggests that the source is marginally resolved in the convolved 15\,GHz image, we do not believe this to be significant given that it is unresolved in the native resolution 15\,GHz image and at all other frequencies in both the native or beam-matched images.  
If we were to instead treat the fitting result to be unresolved, the total flux density obtained by the geometric mean of the peak brightness and integrated flux density is still within $2 \sigma$ of the integrated value (i.e., $163.\pm 11.\,\mu$Jy), and thus would not affect our results.

\subsection{Modeling the Radio Spectrum \label{sec:spec}}
We fit the radio data using three separate components: thermal dust emission, thermal (free-free) radio emission, and a spinning dust component.  
We do not include a component for non-thermal synchrotron emission as the data do not suggest the presence of a steep spectral component between 3 and 15\,GHz, thus the inclusion of such a component would be completely unconstrained.  
The lack of a non-thermal component is also suggested by an attempt to split the 3\,GHz data into two sub-band images centered at 2.5 and 3.5\,GHz, for which we measure a spectral index that is consistent with optically thin free-free emission (i.e., $\alpha \approx 0.06 \pm 0.41$).

Since the shape and amplitude of the far-infrared emission is not directly constrained by the data in hand, we assume a typical infrared spectral energy distribution using the \citet{dh02} templates and assign an infrared (IR; $8-1000\,\mu$m) flux based on the measured 3\,GHz flux density assuming it arises completely from free-free emission.  
This is done by equating Equations 6 and 15 in \citet{ejm12b}. 
The infrared flux is then taken to be $F_{\rm IR} \simeq 2.5\times10^{-14}\,{\rm W\,m^{-2}}$, corresponding to an infrared luminosity of $L_{\rm IR} \simeq 1.1\times10^{8}\,L_{\sun}$.  
We believe that the accuracy of the assumed infrared flux not better than a factor of $\gtrsim 2$, which is the scatter measured between the 33\,GHz and IR flux on larger, $\approx$\,kpc-scale regions in the SFRS galaxies \citep{ejm12b}.

For the free-free emission, we first assume a component that scales as $S_{\nu}^{\rm T} \propto \nu^{\alpha_{\rm T}}$, where $\alpha_{\rm T}=-0.1$ is the thermal spectral index. 
We do not include the possibility for optically-thick free-free emission given that the spectrum does not continue to fall, but rather flattens, with decreasing frequency between 15 and 3\,GHz. 
For the spinning dust component, we use the models of \citet{ahd09, sah11} and the associated publicly available code {\sc spdust2}\footnote{\url{http://cosmo.nyu.edu/yacine/spdust/spdust.html}} assuming a range of warm neutral medium (WNM; $x_{\rm H} \approx 0.1$), warm ionized medium (WIM; $x_{\rm H}\approx 0.99$), Molecular Cloud (MC; $x_{\rm H}\approx 0$), and Dark Cloud (DC; $x_{\rm H}\approx 0$) conditions as defined by \citet{dl98b}, where $x_{\rm H}$ is the ionization fraction.  

For each idealized ISM environment, we varied the gas density ($n_{\rm ISM} = 0.1, 0.15, 0.3, 0.5, 1, 1.25, 1.5$\,cm$^{-3}$) and gas kinetic temperature ($T = 0.1,0.3,0.5,0.7,1\times10^{4}$\,K).  
For MC and DC cases, we used different ranges for $n_{\rm ISM} =0.01,0.05,1,2, 5\times10^{4}$\,cm$^{-3}$ and $T =1, 2, 5, 7, 10, 15, 20$\,K.    
The amplitude of the free-free emission and spinning dust components were then varied to best fit all radio points, using a standard $\chi^{2}$ minimization procedure.  
The fit to the radio spectrum of NGC\,4725\,B is shown in Figure \ref{fig:spec}, along with the relative contribution of each component.  


\section{Results}
In the following section we describe the results from the radio imaging and photometry.  
To better elucidate the source of the emission, we examine the characteristics of NGC\,4725\,B at other wavelengths.
We first establish whether the source is actually located in NGC\,4725, or is a background $\mu$Jy radio source.  
We conclude that the source is most likely associated with NGC\,4725, and that spectrum is best explained by including a significant contribution from AME.

\subsection{Background Radio Source}

Figure \ref{fig:radpan} presents colorscale images at all observed radio frequencies: 3, 15, 33, and 44\,GHz.  
We show both the original observation at 33\,GHz taken 2013 March and the most recent observation obtained via our DDT proposal 2017 July, the latter being observed in C-configuration, and at significantly higher angular resolution.   
In the first panel a 1\arcsec~scale bar, corresponding to a linear scale of $\approx 60$\,pc at the distance of NGC\,4725, is shown in the bottom left corner.  
The size of the synthesized beam is also shown in the bottom right corner of each panel.  
A 5\arcsec~($\approx 300\,$pc) diameter circle centered on the NGC\,4725\,B is shown in each panel.  
NGC\,4725\,B is clearly detected at all frequencies, and appears to be compact, as its extent is similar to the size of the beam at all frequencies.    

In Figure \ref{fig:spec} the spectrum of NGC\,4725\,B is plotted, illustrating that the flux density increases between 3 and 15\,GHz, then sharply rises and peaks near 33\,GHz.  
The fact that the source appears to peak at 33\,GHz already suggests that it is not likely a background radio galaxy, given that there are no known sub-mJy radio sources that peak at such high frequencies.  
Taking the 3 sources in \citet{hancock10} detected in the ATCA 20\,GHz \citep[AT20G;][]{at20} survey with spectra that peak near $\approx$30\,GHz, we use their luminosities to estimate the corresponding redshifts at which we would expect to detect such a faint radio source.  
With a 33\,GHz flux density of $S_{\rm 33\,GHz} \approx 402.\,\mu$Jy, NGC\,4725\,B would need to be located at $z \gtrsim 50$ and peak in the rest-frame at $\gtrsim 1.67$\,THz to have the same intrinsic luminosities of these 3 sources, which seems completely implausible.  

Given that the flux density limit of AT20G ($4\sigma \approx 40$\,mJy) is significantly larger than for the full 33\,GHz SFRS imaging \citep[$4\sigma \approx 75\mu$Jy][]{ejm18a}, we attempt to estimate the likelihood of detecting a fainter population of inverted spectrum sources reported in \citet{hancock10}.  
We do this by taking the fraction of extragalactic AT20G sources having rising spectra between 5 and 20\,GHz reported in \citet[][i.e., 5.8\%]{massardi11}, then correct for the difference in flux density limits using the 28.5\,GHz stacked {\it Planck} source counts of \citet{ejm18b}.  
This results in an expectation of $\lesssim 0.2$ faint, inverted spectrum sources detected in the full 33\,GHz SFRS imaging.  
Thus, we would need to map $\gtrsim 5\times$ the area covered by the entire SFRS at 33\,GHz to detect a single source.  


\subsubsection{Identification of Optical and Infrared Counterparts}

In Figure \ref{fig:ancpan} we show colorscale optical ($BVRI$) and {\it Spitzer}/IRAC (3.6, 4.5, 5.8, and 8\,$\mu$m) images that were observed as part of SINGS \citep{rck03}.  
A 5\arcsec~($\approx 300\,$pc) diameter circle centered on the NGC\,4725\,B is shown in each panel, which corresponds to the same circle shown in Figure \ref{fig:radpan} to allow for comparison between all panels.  
As in Figure \ref{fig:radpan}, a 1\arcsec~scale bar, corresponding to a linear scale of $\approx 60$\,pc at the distance of NGC\,4725, is shown in the bottom left corner of the first panel.    

There is no obvious detection of NGC\,4725\,B in any of the optical bands.  
However, there appears to be a counterpart in each of the IRAC bands, with the source structure increasing in its complexity with increasing wavelength.  
The source appears to be rather distinct at 3.6 and 4.5\,$\mu$m, becomes a bit more extended at 5.8\,$\mu$m, and seems to be part of a larger dust lane that is clearly visible at 8\,$\mu$m.  
Looking at the 3.6\,$\mu$m image, there does appear to be low-level (presumably dust) emission connecting the inner disk to NGC\,4725\,B.  
Consequently, this lends further evidence that NGC\,4725\,B is associated with NGC\,4725 and not a background radio source.

\subsection{An Explosive Transient}

Another possibility for the origin of this source is an exploding transient.  
Looking at its relative brightness as a function of time, we do not believe that this scenario is likely. 
The 3, 15, and original 33\,GHz observations were carried out 2013 November, 2014 November, and 2013 March, respectively. 
If the source were an explosive transient, we would expect that the source brightness would increase or decrease monotonically with time, which is not observed.  
Furthermore, the 33\,GHz observations span a 4\,yr period and do not show evidence for variation.  
If this were an explosive transient, we would expect its brightness to peak earlier at higher frequencies and continue to increase with decreasing frequency over time, which is not the case.

\begin{figure}[t!]
\epsscale{1.25}
\plotone{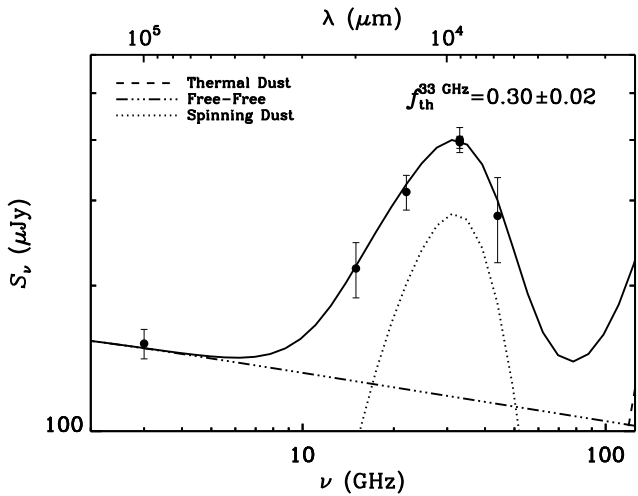}
\caption{
The full radio spectrum of NGC\,4725 shows strong peak at $\approx$33\,GHz.  
The 33\,GHz flux densities measured in the original and re-observed data are both plotted and statistically indistinguishable (see Table \ref{tbl-1}).  
Given the much higher S/N for the flux density measured in the 2013 March 33\,GHz image, this value was used for fitting the spectrum with the various emission components as shown (i.e., free-free, spinning dust, and thermal dust emission).   
The best-fitting spinning dust model is for WNM conditions with $n_{\rm ISM} =1.25$\,cm$^{-3}$ and $T = 1000$\,K, and having a 33\,GHz thermal radio fraction of $f_{\rm th}^{\rm 33\,GHz} = 0.30\pm0.02$.  
 }
\label{fig:spec}
\end{figure} Enuc

\subsection{Anomalous Microwave Emission} 

The most likely origin of the 33\,GHz emission in NGC\,4725\,B appears to be AME associated with spinning dust grains.
In Figure \ref{fig:spec} we fit a spinning dust model (see \S\ref{sec:spec}) to see if such an emission component is in fact plausible.  
Based on a simple $\chi^2$ minimization, the spinning dust component is able to fit the observations extremely well, and appears most consistent with ISM density and gas temperature conditions associated with the WNM (i.e., $n_{\rm ISM} =1.25$\,cm$^{-3}$ and $T = 1000$\,K), similar to the best-fit spinning dust models found for NGC\,6946 \citep{ejm10}.  
The $\chi^2$ values are similarly small for WIM conditions, and only somewhat larger for both the MC and DC conditions.  
Consequently, the exact model is not highly constrained given the spectral sampling of the available data, as there seems to be a range of acceptable values.

The 33\,GHz thermal radio (free-free) fraction is $f_{\rm th}^{\rm 33\,GHz} = 0.30\pm0.02$, similar to the value of $f_{\rm th}^{\rm 33\,GHz} \approx 0.43$ measured for NGC\,6946 Enuc.\,4 \citep{bh15}. 
Given the distances to NGC\,4725 and NGC\,6946, the 33\,GHz spectral luminosity of the spinning dust emission is remarkably similar for both sources, being $L_{\rm 33\,GHz}^{\rm AME} \approx 4.7$ and $4.5\times10^{18}$\,W\,Hz$^{-1}$, respectively.
As shown in Figure \ref{fig:ancpan}, the source is clearly detected in each of the mid-infrared IRAC bands, and appears to be extremely compact.  
The fact that the source is so strongly detected in the lowest wavelength bands (3.6 and 4.5\,$\mu$m) suggests a significant amount of small hot dust grains associated with this source, consistent with the spinning dust interpretation.

\section{Discussion}
In this section we discuss NGC\,4725\,B in the context of other known detections of AME in search for any unifying characteristics among them.  

\subsection{Environmental Conditions}
The emissivity of the best-fit spinning dust model corresponds to a hydrogen column density of ${\rm N_{H}} \approx 1.5\times10^{23}\,{\rm cm^{-2}}$, which is a rather large value and suggests a very dense region. This is true regardless of which ISM condition is used for the fit.  
However, looking at the archival BIMA Survey of Nearby Galaxies \citep[SONG,][]{bimasong} CO ($J=1\rightarrow0$) map of NGC\,4725, there is no obvious counterpart detected, which may in part be due to its much lower ($\theta_{1/2} \approx 7\arcsec$) angular resolution.  
There does seem to be some associated CO ($J=2\rightarrow1$) emission \citep{heracles09}, however the angular resolution of the IRAM data are even lower resolution ($\theta_{1/2} \approx 13\farcs4$), making it difficult to associate the CO emission with NGC\,4725\,B. 

One possibility then is that this is a nascent star-forming region where massive stars are still highly enshrouded by their natal cocoons of gas and dust \citep[e.g.,][]{hr03}.  
The lack of any apparent non-thermal synchrotron component suggests that these massive stars are extremely young ($\lesssim 3$\,Myr) with very few having exploded as supernovae.  
Assuming again that all of the 3\,GHz flux density arises from free-free emission, we convert this to an ionizing photon rate of $Q(H^{0}) \approx 1.8\times10^{51}\,$s$^{-1}$ \citep[][Equation 10]{ejm11b}, which equates to $\approx 160$ O7 stars, where $Q(H^{0}) \approx 10^{49}\,$s$^{-1}$ per effective O7 star \citep{sternberg03}.  

This scenario may be consistent with the occurrence of the AME observed for NGC\,6946 Enuc.\,4, which sits on the rim of an H{\sc i} bubble (seen also at 8\,$\mu$m) in the arm of the disk and is similarly compact.  
The nearby massive star-forming region is able to apply a strong, incident radiation field that can heat a large accumulation of dust located within the rim of the gas/dust shell in which the H{\sc ii} region resides.  


\subsection{Relation Between AME and Thermal Dust Emission}
In the Galaxy, AME appears to be a ubiquitous feature of the dust emission, with the AME having a roughly linear dependence on various tracers of dust emission (e.g., 100\,$\mu$m emission, dust radiance) across the sky \citep[e.g.,][]{Davies2006,Planck2016-XXV,bh16}. A key question, then, is whether these extragalactic regions have unusually strong far-infrared dust emission relative to their radio brightness, or whether the AME is enhanced in these regions without a corresponding increase in the thermal dust emission.

In the sample of extranuclear regions presented by \citet{ejm18a}, both NGC\,4725\,B and NGC\,6946 Enuc.\,4 are clear outliers in $S_{33\,{\rm GHz}}/f_{24\,\mu{\rm m}}$, differing from the other regions by roughly an order of magnitude. A constant ratio between the 33\,GHz AME flux density and the 24\,$\mu$m flux density for all regions is not tenable regardless of the 33\,GHz AME fraction in the sources without an AME detection. Since NGC\,6946 Enuc.\,4 is not unusual in its ratio of 24\,$\mu$m emission to total dust emission \citep{ejm12b}, it appears that these two AME sources have unusually strong AME per thermal dust emission relative to other star-forming extranuclear regions.

The second question of interest is how the AME per thermal dust emission in these sources compares to what is observed in the Galaxy. Comparing the diffuse Galactic AME observed by {\it Planck} to the far-infrared dust emission, \citet{bh16} found a linear relation between the dust radiance $\mathcal{R}$ and the 30\,GHz AME intensity with a conversion factor of $6420\pm1210$\,MJy\,sr$^{-1}$/(W\,m$^{-2}$\,sr$^{-1}$). 

Unfortunately, the available far-infrared data for NGC\,4725\,B are of insufficient resolution to make robust determinations of its thermal dust emission. 
Although we cannot measure the IR flux from NGC\,4725\,B directly, if we take $F_{\rm IR} \simeq 2.5\times10^{-14}$\,W\,m$^{-2}$ as estimated from the 3\,GHz free-free emission in Section~\ref{sec:spec}, we would expect $\sim 160$\,$\mu$Jy of AME at 30\,GHz from the \citet{bh16} relation versus the $\sim 280$\,$\mu$Jy observed at 33\,GHz. We caution that taking these scaling relations at face value {\it generically} implies a ratio between the 30\,GHz AME and free-free flux densities of 1.3. While consistent with NGC\,4725\,B and NGC\,6946 Enuc.\,4, this level of AME does not appear consistent with the other star-forming regions in the SFRS sample \citep{ejm18a}. 

NGC\,6946 Enuc.\,4 has a ratio of 30\,GHz AME to IR emission of $\simeq 1.8\times10^4$\,MJy/(W\,m$^{-2})$ \citep{bh15}, a factor of $\sim2.8$ larger than expected from the Galactic relation. Similarly, its ratio of 30\,GHz AME to 100\,$\mu$m flux density of $\simeq6\times10^{-4}$ \citep{bh15} is a factor of $\sim2$ larger than the Galactic mean \citep{Planck_Int_XV,bh16}. Thus the AME in NGC\,6946 Enuc.\,4 appears to be a factor of 2--3 stronger than expected from its thermal dust emission relative to what is observed in the Galaxy.

It is of course possible that NGC\,4725\,B and NGC\,6946 Enuc.\,4 represent the extreme that is most favorable for AME detection, and the remaining regions still harbor AME but at a somewhat lower level. A clear test of the ubiquity of AME as a component of the dust emission would be a direct measurement of the far-infrared fluxes in these regions and comparison with upper limits on their $\sim 30$\,GHz AME. Isolating the factors governing the level of AME in these regions will lend insight into the physical mechanisms powering the AME as well as the nature of its carrier(s).

In addition to IR fluxes and more stringent limits on AME at $\sim 30$\,GHz, spectroscopy of the 3.3 and 3.4\,$\mu$m PAH features in both NGC\,4725\,B and NGC\,6946 Enuc.\,4 would test the association between the AME and the smallest grains and perhaps provide clues to the molecular nature of the AME carrier(s). Such observations should be feasible in the near future with the Near-Infrared Spectrograph aboard the James Webb Space Telescope.



\section{Summary}
Similar to the detection of AME within the disk of NGC\,6946, we find significant AME in only a single, discrete region of NGC\,4725.  
These results suggest that radio observations from external galaxies may contain appreciable amounts of AME, complicating the typically assumed picture that the extragalactic radio emission is only comprised of non-thermal and free-free emission.  
While certainly true for discrete regions, it remains less clear how much of an effect this component has for interpreting globally integrated measurements of galaxies.  
For instance, \citet{peel11} did not find a strong excess in the WMAP/{\it Planck} spectra for three nearby bright galaxies, M\,82, NGC\,253, and NGC\,4945, suggesting that AME may not affect simple two-component modeling of extragalactic sources on a global scale. 
However, this may not be the case at high redshift where galaxies have been modeled to have significantly different grain properties \citep[e.g.,][]{maiolino04, dap10}.  
For instance, dust grain properties of low metallicity populations at high redshift are found to be similar to that of the Small Magellanic Cloud \citep[e.g.,][]{nr15,nr18}, a system that appears to contain a significant amount of AME arising from a combination of spinning and magnetized nano-grains \citep{dh12}.
More work is clearly needed to understand the physical underpinnings that drive AME within external galaxies on all physical scales. 

\acknowledgements
We would like to thank the anonymous referee for very useful comments that helped to improve the content and presentation of this paper.  
E.J.M thanks J.J. Condon and B.T. Draine for useful discussions that helped improve the paper.  
The National Radio Astronomy Observatory is a facility of the National Science Foundation operated under cooperative agreement by Associated Universities, Inc.
This research was carried out in part at the Jet Propulsion Laboratory, California Institute of Technology, under a contract with the National Aeronautics and Space Administration.

\bibliography{aph.bbl}

\end{document}